\def\sun{$_\odot$~}
\def\etal{{\it et al.}}
\documentstyle[12pt]{article}
\def\beginapjbib{\begingroup \section*{\large \bf References}
         \parskip=.5ex plus 1.0pt
         \def\bibitem{\par \noindent \hangindent\parindent
                \hangafter=1}}
\def\endapjbib{\par \endgroup}
\def\beginfig{\begingroup \section*{\large \bf Figure Captions}
         \parskip=.5ex plus 1.0pt
         \def\figitim{\par \noindent \hangindent\parindent
                \hangafter=1}}
\def\endfig{\par \endgroup}
\begin{document}
\begin{titlepage}
\pagestyle{empty}
\baselineskip=21pt
\rightline{UMN-TH-1260/94}
\rightline{astro-ph/9408062}
\rightline{July 1994}
\vskip .2in
\begin{center}
{\large{\bf The Deuterium Abundance and Nucleocosmochronology}} \end{center}
\vskip .1in
\begin{center}

Sean Thomas Scully and  Keith A. Olive

{\it School of Physics and Astronomy, University of Minnesota}

{\it Minneapolis, MN 55455, USA}

\vskip .1in

\end{center}
\vskip .5in
\centerline{ {\bf Abstract} }
\baselineskip=18pt
We examine galactic chemical evolution
models which reproduce the present-day and pre-solar values of deuterium
starting with a primordial value which is consistent with
a baryon-to-photon ratio of $3 \times 10^{-10}$.
We consider various galactic chemical evolution models to determine
the viability of significant deuterium destruction
and which provide a consistent age of the galaxy at the time of
the formation of
the solar system and consequently its present day age from
nuclear chronometers.
These models generally require some amount of infall
which we take with rates proportional to the gas mass as
well as exponentially
decreasing rates and some initial
disk enrichment which we limit to the range of 0\% to 30\%.
We present those models which give
the observed pre-solar value and present-day value of D/H
and which lead to a
present-day gas fraction of $\sigma = .05\quad - \quad .2$.
These models result in a braod range for the age of galaxy between $9.8 - 21.6$
Gyrs.

\noindent
\end{titlepage}
\baselineskip=18pt
\def\la{~\mbox{\raisebox{-.7ex}{$\stackrel{<}{\sim}$}}~}
\def\ga{~\mbox{\raisebox{-.7ex}{$\stackrel{>}{\sim}$}}~}






%

%

\section{Introduction}
One of the successes of the standard big bang
model is its ability to reproduce the
observed abundances of the light elements. In two cases, $^4 He$ and $^7Li$,
the primordial value can be directly determined from observation.
The primordial value of $^4 He$
is inferred from low metallicity HII regions
(see e.g. Pagel et al. 1992; Skillman et al. 1994; Olive \& Steigman 1994).
The consistent abundance of $^7Li$ observed in old halo dwarfs is generally
assumed to
be the primordial value (see eg. Spite \& Spite, 1993).
The primordial abundance of $D$,
however, is much
harder to test observationally since $D$ is destroyed in stars.
A pre-solar value determined
from meteorites, solar wind studies, and the Jovian atmosphere
 (see e.g. Geiss 1993)
and a interstellar value determined
from Space
Telescope observations (Linsky et al. 1992) are all that are known.
There is a reported detection of $D$ in a high redshift,
low metallicity quasar absorption system (Songaila et al. 1994;
Carswell et al. 1994) with an abundance which may be the primordial one.
Due to the still some what preliminary status of this observation and
the fact that it can also be interpreted as a $H$ detection
in which the absorber
is displaced in velocity by 80 km s$^{-1}$ with respect to the quasar
(see also Steigman 1994; Linsky 1994),
we do not fix the primordial abundance with that value (we will
comment further on this possibility below).
The primordial value must therefore be explored
in the context of chemical evolution
models and the observed pre-solar and interstellar values (Audouze and
Tinsley 1974).

Several investigations have explored the relationship between chemical
evolution and the primordial $D$ abundance.
(Vidal-Madjar \&
Gry, 1984; Clayton 1985; Delbourgo-Salvador et al., 1985; Vangioni-Flam and
Audouze, 1988).  It is clear that the degree
to which $D$ is destroyed will be very sensitive to the
chemical evolution model.
 More recently, Steigman and Tosi (1992) explored the acceptable
range for the primordial $D$ abundance without overproduction of $^3He$
with a given set of chemical
evolution models (Tosi, 1988). They found that typically deuterium was
destroyed by no more than a factor of $\sim 2$, leading to the conclusion
that the baryon-to-photon ratio, $\eta$ was relatively high $\eta \ga
4 \times 10^{-10}$.
A different approach was taken by
Vangioni-Flam, Olive, \& Prantzos (1994), (hereafter VOP).  They considered a
wider range of chemical
evolution models starting with a primordial value of $D$ of $7.5 \times
10^{-5}$ consistent with a baryon-to-photon ratio of $3 \times 10^{-10}$
 a value which is more central with respect to $^4He$ and $^7Li$
consistency.
They explored various initial mass functions (IMFs) and star
formation rates (SFRs) to find chemical histories which could reproduce
a $D$ destruction factor of $\sim 5$ to match the presolar value
of 2.6 $\times 10^{-5}$ (Geiss 1993) and the present ISM
value of 1.65 $\times 10^{-5}$ (Linsky et al. 1992). They did not, however,
consider infall in their
models and argued against infall since the observations indicate that the
$D$ abundance has decreased since the pre-solar value. The impact of infall
on these types of models has not formally been tested.

A further complication lies in determining the time at which to compare the
observed abundances with the results from chemical evolution models.
This complication is due to the
large uncertainty in the age of the galaxy.
The age of
the galaxy and pre-solar epochs can be constrained, however,
in chemical evolution
models which consider,
the ratios of long-lived radioactive
nuclei such as $^{232}Th$, $^{235}U$, and $^{238}U$ (Cowan et al. 1987).
The advantage to including this constraint in with chemical
evolution models is two-fold. First, only models which can reproduce the
ratios observed for these elements are considered viable.
Secondly, these models give
times for the formation of the solar system and today at which direct
comparisons to observations can be made.

We will consider the evolution of $D$ using the nuclear chronometer
constraint; that is, we will only consider models which are capable
of producing the correct ratios of $^{232}Th$, $^{235}U$ to $^{238}U$
{\it at the same time}. Solar system studies indicate a pre-solar $D$ abundance
(Geiss
1993) in the range of $(2.6 \pm 1.0) \times 10^{-5}$. The $D$ abundance has
also been determined in the ISM.  Lyman absorption spectra in nearby stars
indicate a present-day $D$ abundance of $(.5 - 2.0) \times 10^{-5}$
(see e.g. Vidal-Madjar 1991, Ferlet 1992). A more
high precision measurement of $D/H$ was determined recently using the HST
(Linsky et al. 1992) which indicates a $D$ abundance of $1.65^{+.07}_{-.18}$.
For this work, we will adopt
the HST measurement for the present-day value.

Our goal therefore
is to explore chemical evolution models which can account for the
deuterium abundances observed at the pre-solar epoch and today from a
primordial value of $\sim 7.5 \times 10^{-5}$ constrained by nuclear
chronometers. It has been argued previously (Reeves 1991; Reeves 1994)
that because of the uranium chronometers, the average
galactic nuclear activity today is within a factor of two of its average
value over the lifetime of the galaxy and therefore may limit the degree
to which $D$ is astrated.
 To test this hypothesis, we consider various initial mass functions (IMFs),
star
formation rates (SFRs) and infall rates which in fact
can achieve this goal. We also demand
 that the models reproduce a gas fraction, $\sigma$, in
the range of .05 to .2 and require a maximum initial enrichment of 30\%
with respect to present abundances of
heavy elements (for a review see Rana, 1991).  Finally, we will consider the
oxygen abundances predicted from such models as a gauge for heavy element
production.

\section{Chemical Evolution Models}

\subsection{General Equations}

In order to examine
chemical histories which lead to the observed abundances of $D$,
we consider a variety of numerical chemical evolution models.  We employ a
chemical evolution code which avoids the instantaneous recycling
approximation (IRA), that is, the actual stellar lifetimes are considered in
the calculation.  We consider a variety of IMFs, SFRs, and infall rates.  For
simplicity, we
neglect radial gas flows within the disk for our models, and we do not
assume that the galactic disk is homogeneous, but restrict our
calculation to the solar neighborhood. We make no
assumptions on the age of the disk.  The age is constrained in our model by
the nuclear chronometers.

Age determinations, which consider the r-process chronometer pairs
$^{235}U/^{238}U$ and $^{232}Th/^{238}U$ have been considered by a number of
authors (for review see Cowan et al., 1991).
These calculations start from the
basic equation for gas mass within the galactic disk
$${{dM_G}\over{dt}} = e(t) -\psi(t) + f(t), \eqno (1)$$
where $e(t)$ is the ejection rate,
$$e(t) = \int\limits_{m(t)}^{m_{upp}}\!{(m-m_R)\psi(t-\tau_m)\phi(m)}\,dm.
\eqno
(2)$$
In these equations,
$f(t)$ is the infall rate onto the disk, $\phi(m)$ is the IMF, and
$\psi(t)$
is the SFR.  $m_{upp}$ is the upper mass limit on $\phi(m)$.
$\tau_m$ is the lifetime of a star of mass $m$,
and $m_R$ is the remnant mass of a star of mass $m$.  $m(t)$ is is the mass
of a star which at time $t$ is returning gas back into the ISM.
$m_R$ is taken to be (Iben \& Tutukov, 1984),
$$m_R = .11m + .45  M_{\odot} \quad m \leq 6.8  M_{\odot}$$
$$m_R = 1.5  M_{\odot} \quad m > 6.8  M_{\odot}. \eqno (3)$$
Stellar lifetimes are determined as in Scalo (1986).

Equation (1) can be extended to determine the rate of change in the number of
nuclear species $A$,
$${{dN_A}\over{dt}} = P_A\psi(t) - {{\psi(t)N_A}\over M_G} + {{e(t)N_A}\over
M_G} + {f\over M_G}{Z_f\over Z}N_A - \lambda_A N_A. \eqno (4)$$
In this equation, $P_A$ is the number of newly synthesized nuclei of species
$A$
per unit mass going into star formation.  The relative production ratios,
$P_{232}/P_{238}$, and $P_{235}/P_{238}$, are taken to be $P_{232}/P_{238}$
= 1.60, and $P_{235}/P_{238}$ = 1.16,
adopted from Cowan et al. (1987).
$\lambda_A$ is the rate of decay of nuclear species $A$.  For
simplicity, the amount of species ``A" which decays while locked up in
stars has not been considered. This is a good approximation since
only the largest mass stars will significantly
contribute to the amount of ``A" produced and returned to the ISM and these
stars have lifetimes which are short compared with the decay rate of
$^{232}Th$, $^{235}U$, and $^{238}U$.  The
decay rates used in the calculation are based on the half-lives
of the respective parent nuclei. For this work, we have adopted half lives
of $\tau_{232}
= 1.41 \times 10^{10}$ yrs, $\tau_{235}= 7.04 \times 10^{8}$ yrs, and
$\tau_{238}= 4.46 \times 10^{9}$ yrs.
We have also assumed the
metallicity of the infall gas, $Z_f$, to be zero.

Equation (4) can now be solved numerically simultaneously with equation
(1). As a boundary condition for equation (4), $N_A(0)$ is allowed to be
non-zero to satisfy observational constraints on the
metallicity distribution and age-metallicity relation of stars
(Pagel and Patchett 1975; Twarog 1980). This is equivalent to assuming an
initial burst of star formation. $N_A(0)$ is a free parameter in our
calculation and the value of $S_0$, the initial disk enrichment,
that results from the choice of $N_A(0)$ is also calculated.  Initial disk
enrichments of $S_0 \la .3$ are considered acceptable.  This corresponds to
observational constraints necessary to account for the observed metallicities
(Lynden-Bell 1975).

The final consideration is the abundance of $D$.  We assume no galactic
sources of $D$.  Furthermore, we consider $D$ to be totally astrated within
stars.  Under these assumptions, the rate of change of $D$ in the disk can
be determined by extending equation (1),
$${d\over {dt}}(DM_G) = -\psi(t)D + f(t)D_f, \eqno (5)$$
where $D_f$ is the mass fraction of $D$ in the infall gas
which we assume to be primordial.  By substitution
of equation (1) into this equation, it can be rewritten in the form
$${dD\over {dt}} = -e(t)D + f(t)(D_f-D). \eqno (6)$$
This equation is then solved simultaneously with equations (4) and (1).

\subsection {Model Parameters}

The models are divided into three groups based on their SFRs and are
closely related to the models studied in VOP.
In group I, the
star formation rate is assumed to be proportional to the gas mass ($\psi(t)
= \nu M_G$).
Group II considers exponential star formation rates ($\psi(t) = \nu
e^{-t/\tau}$).
We explore decay constants, $\tau$, with values $\tau =
\infty, 10, 8, 5, 3 \quad Gyrs^{-1}$.
Finally, group III models assume a star formation rate proportional to the
gas mass
squared ($\psi(t) = \nu {M_G}^2$). We allow the constant of proportionality
in each of these cases to be a free parameter.
For most models a simple power-law IMF has been chosen. We do consider the
effects of varying the IMF.
In what we label group Ib, we consider an IMF as determined by Tinsley (1980).
The effect of varying the IMF is also considered for group II models.  Group
IIb models employ the IMF given by Scalo (1986) and group IId models are
based on Tinsley (1980) with an exponential SFR. The IMF is
normalized
$$\int\limits_{m_{low}}^{m_{upp}}\!{m\phi(m)}\,dm = 1, \eqno (7)$$
where $m_{low}$ and $m_{upp}$ correspond to the lower and upper mass limits on
stars.  The values of $m_{low}$ = .4 and $m_{upp}$ = 100 have been chosen
for this calculation except in the case of the Tinsley (1980) and Scalo
(1986)
IMFs where  the value $m_{low}$ = .1 is used.
In VOP, a power-law IMF with  $\phi(m) \sim m^{-2.7}$ was used and was found
to be sufficient for obtaining the required $D$ destruction. When we impose the
chronometer constraint, we find that in many cases not enough $D$ was destroyed
and/or the gas mass fraction tended to be too low. That is, depending on the
assumed SFR, solutions which yield consistent ages, did not always destroy
enough $D$ and in some cases (eg. models IIa) very few or no solutions were
found that destroyed
$D$ by a total factor of 5 with a sufficiently high gas mass fraction. By
choosing a flatter IMF, more massive stars which destroy
$D$ are produced. Here a  power-law IMF with
$\phi(m)
\sim m^{-2.35}$, has been assumed for most of the models.
For model III,  an even flatter power law ( $\phi(m) \sim
m^{-2.05}$) was required  models to destroy enough $D$ to match
observations.

Two commonly chosen forms for the infall rate are examined in our calculation.
The first of these is an infall rate proportional to the gas mass ($\sim
\mu M_G$).  Though there is no physical reason for assuming this kind of infall
rate, under certain assumptions, an
infall rate proportional to $M_G$ can lead to analytic
solutions (see e.g. Clayton
1984,1985).  We also consider infall rates which are exponential ($\sim \mu
e^{-t/\tau}$) in models Ic and IIc.  Decay constants of $\tau =
\infty, 10, 8,$ and $3 \quad Gyrs^{-1}$ are examined.
This type of model was previously examined by Tosi (1988).
 A summary of all the models and the various parameters
chosen can be found in table (1). We have imposed a cut-off on the total disk
mass $M_T$ to be no larger than three times the initial disk mass.

In addition to the constraints outlined above, the abundance of $^{16}O$ will
be
determined for each of the model types where the yields of Woosley (1993) have
been adopted.  For models which overproduce oxygen, the value of
$m_{upp}$, the upper mass limit of stars which supernova, is lowered.
The impact on the evolution of $D$ by
lowering this value to reproduce the $^{16}O$ abundance will
be discussed below.

\subsection{Model Calculation}

The calculation is divided into two parts.  In the first,
we search for matches for the ratios of the nuclear
chronometers $^{232}Th/^{238}U$ and $^{235}U/^{238}U$ determined from
meteoritic abundances.  The abundance ratios used here are $^{232}Th/^{238}U$ =
2.32 and $^{235}U/^{238}U$ = .317 (Anders and Ebihara 1982) which have an
error of 5 $-$ 10\char37.  In our calculation, we have
allowed for an error on these values of $\pm 5\char37$.  We then look for
solutions within this range allowing $N_A(0)$, $\nu$ and $\mu$ to vary as free
parameters. Chemical histories which produce matches for both ratios
at the same time are then considered
potential solutions.  The model
time at which the ratios match is then considered
the pre-solar epoch and subsequently 4.6 Gyrs later, today.
Next, the $D$
abundance is calculated for models which reproduce these ratios.  We test
for
models which can reproduce the $D$ abundance at the times designated as
the formation of the solar system  and ``today"
and which result in a gas mass fraction
$\sigma$ within the range of $.05-.2$.
We examine those models in the next section.

\section{Results}

Most of the model types explored were able to produce chemical evolution
scenarios compatible with the observations of the abundance of $D$. We begin
by considering the general characteristics that resulted from group I models.
Group Ia models which reproduce the
observations of $D$ lead to a low of a gas fraction
($\sigma \sim .05 - .08$) which is only marginally in agreement with
observations. A higher gas mass fraction can be obtained by further flattening
the IMF. For example, when $\phi(m) \sim m^{-2.7}$, solutions give a value of
$\sigma$ no larger than $\sim 0.05$, while for $\phi(m) \sim m^{-2.05}$ we find
solution with $\sigma$ as high as 0.2 for model Ia. It is also important to
note that our calculational scheme was not designed to maximize the gas mass
fraction.  Instead our solutions minimize the error in the chronometer
abundance ratios.  When this condition is relaxed (though still enforcing
ratios within 5\% of their observed abundances) significantly higher gas mass
fractions are found.

 In group Ia models, as much as 50\% to 70\% of the initial disk
mass falls onto the disk over the galaxy's evolution in these models. In
general, the higher the star formation rate, the higher the resulting disk mass
(see figure 1a).  The solutions shown here have been restricted by the
requirement that the present $D$ abundance match the ISM measurement to
within 3 \%. With a higher star formation rate, more
$D$ is destroyed.  The only source of $D$ is the infall gas.  As a result, a
larger amount of infall is required to reproduce the observations of $D$ for a
higher star formation rate.

Group Ib model solutions closely resemble those of group Ia.  They do
result in a slightly higher gas fraction than those
of Ia ($\sigma \la .09$) which is in better agreement with observations.
The IMF for these models is flatter for the lower mass stars. As
a result, less mass is
locked up in these stars leading to a larger return to the ISM.  This allows
for more $D$ destruction  at a higher resulting $\sigma$.
These models require even more
infall onto the disk than group Ia; up to 90\% of the
initial disk mass is
required to fall onto the disk to achieve solutions.

No viable $D$ evolution was found for group Ica models
(those with a constant infall rate). These models result
in a very low $\sigma$ ($\sigma \sim .03$) before enough $D$ is destroyed to
match observational constraints.
Group Icb models can reproduce $D$ observations with very little
infall (only 27$-$39\% of the initial disk mass), although
they do result in a $\sigma$ on the low
side ($\sigma \sim .05 - .07$). Steeper exponents for the infall rate
(groups Icc-Ice) can also produce viable $D$ evolution scenarios with higher
gas fractions at the expense of larger amounts of infall required.
In order to achieve a gas fraction of $\sigma \sim .1$,
at least 70 $-$ 90\% of
the initial disk mass is required to fall onto the
disk for these model types.  Unlike groups Ia and Ib,  the total mass that
results for these model types decreases as a function of the star
formation rate (as seen in figure 1a).   Note however, that in models such as
Icc, there are solutions in which $M_T$ is greater than 3.  (They are found at
$\nu \sim 0.7$, where our imposed cut-off is visible.)

Group IIa models which can reproduce the observations of $D$ lead to a gas
fractions throughout the range of $\sigma \sim .05 - .20$.
Similar to group I types, the higher the resultant
SFR, the higher the resultant
total mass.  Figure 1b clearly demonstrates this trend for group II
models. In addition, as can be seen from figure (2) that in
general, as the exponent of the star formation rate steepens, we find
a lower resulting disk mass and $\sigma$ which themselves are correlated
within each model type.

The amount of
initial disk enrichment required changes as a function of the decay constant
of the star formation rate.
Group IIaa solutions ($\tau = \infty$) result in initial disk
enrichments of 10 $-$ 30\char37.  In general, the steeper the exponent of the
SFR, the less initial disk enrichment is required.  Group IIae solutions for
example all require less than 2\% initial disk enrichment.
Figures 3a and 3b show the range of initial
enrichments for solutions as a function of star formation rate.
Models (except for those with exponential infall rates) with the lowest infall
rates and consequently lowest star formation rates lead to the lowest required
initial disk enrichments. This behavior can be credited to the age at which
the pre-solar epoch occurs for these solutions.
Solutions which have low initial enrichments correspond to later pre-solar
epochs. Star formation and infall occur for a longer period of time before the
pre-solar epoch and thus proceed at lower rates but require less of an initial
disk enrichment to obtain a solution. Figures 4a and 4b show the relationship
between the initial disk enrichment,
$S_0$, and the pre-solar epoch.  The trend of the lower enrichment-later
pre-solar epoch is
demonstrated for the group I and IIa models. Other model groups produce similar
results (this is true for all models including those with exponential infall
rates). From these figures, one sees clearly the range of ages for the galaxy
at the pre-solar epoch that we obtain in our solutions.

Figures 5a and 5b show a comparison of the
evolution of $D/H$ for model groups Ia, Ib, Icb and
IIab and their resulting gas fraction.  The types  are chosen which
result in a realistic evolution for $D/H$ and satisfy the gas fraction,
disk enrichment and nuclear chronometers constraints.
For comparison, solutions were chosen which had equal
pre-solar ages.
All of these models fall within the errors of both the pre-solar
observations and the ISM value.  There is apparently
very little effect on the $D/H$ evolution due to the IMF's tested, as these two
curves nearly overlap.  There is however an effect on the $^{16}O$ abundance
based on the choice of IMF which will be discussed below.
These solutions are not unique. Other
combinations of IMFs, SFRs, infall rates, and resulting pre-solar epochs
could produce viable $D/H$ evolutions as well.
It should be noted that the shape of a solution for the $D/H$
evolution for a particular
model type does not change for differing pre-solar epochs.  The general
character of each model type remains fixed.

In all of the cases which produce a realistic $D/H$ evolution, the $^{16}O$
abundance is overproduced.
Adjusting the upper mass
limit of stars which supernova ($m_{upp}$) for these cases can lower the
$^{16}O$ abundance to its observed levels. Such an adjustment is frequently
necessary when considering exponentially decreasing SFRs (Larson, 1986; Olive,
Thieleman, \& Truran 1987; VOP). The effect of lowering $m_{upp}$ on the $D$
abundance was also considered here. Group I models required lowering the value
of
$m_{upp}$ to $\sim 20$ M\sun in order to match the solar $^{16}O$ abundance.
They suffer virtually no change in their predicted $D/H$
evolution by changing this value.  In contrast, group IIa models are
drastically changed.  $D$ is virtually all destroyed as a result of lowering
$m_{upp}$. This result differs from models which assume the instantaneous
recycling approximation (IRA) (see e.g. VOP).
Changing $m_{upp}$ to a lower value would result in a
lower return fraction.  This would mean that less $D$ would be destroyed
leading to a higher $D$ abundance.  In our models, the infall rate has been
set proportional to the gas mass.  This means that less infall will occur in
a model where less gas is being returned to the ISM due to a lower
$m_{upp}$.  Without this additional source of $D$, the overall effect is to
lower the $D$ abundance.  Group IIb models behaved similarly although the
$^{16}O$ is not overproduced to as great of an extent due to the IMF being
normalized more towards the low end stars.
Group Ia and Ib models have this same assumption.
However, in addition, the star formation rate is proportional to the gas
mass.  Thus with less gas returned, the star formation rate is lower,
leading to less $D$ destruction.  This effect compensates for the depleted
source of $D$ due to a lower infall rate leaving the overall $D$ abundance
virtually unchanged. Group Ic models, which have exponential infall rates,
wind up with an {\it increased} abundance of $D$ as a result of lowering
$m_{upp}$.  In these models, since the infall is independent of the gas
mass, less stellar processing occurs due to the shift towards the low mass
stars while $D$ is still supplied by the infall gas.

Finally, we have considered the effect of changing the $D$ abundance of the
infall gas.  All of the models discussed thus far have considered the $D$
abundance of the infall gas to be primordial.  In general, it is more
difficult to destroy enough $D$ for models which include infall.  Lowering
the abundance of $D$ in the infall gas should, in fact, make it easier to
destroy $D$.  As a test case, group Ia models were rerun assuming an infall
abundance of $D$ to be 50\% primordial and also with no $D$.  Figure 6
is a comparison of the $D$ abundance for a group Ia solution.  $D$
is destroyed by a factor of $\sim 9$ for a model with 50\% of
primordial $D$ abundance in the
infall gas. With no $D$ in the infall,
$D$ is virtually totally destroyed.

\section{Conclusion}

We have shown that $D$ can be destroyed by a factor of $\sim 5$ under a wide
variety of choices of SFRs, IMFs, and infall rates even with the additional
constraint of nuclear chronometers.  It does not seem necessary to constrain
the initial SFR to within a factor of 2 of the
current rate as suggested by Reeves
(1991, 1993) in order to reproduce the ratios of the uranium
chronometers. Though we did find many solutions, we were in fact strongly
constrained by the chronometers.  For example, as in Cowen, Thielemann, \&
Truran (1987), we could not find solutions without infall or without some
initial enrichment.  The necessity of infall was cause for concern with
regard to finding solutions with sufficient $D$ destruction factors (VOP).
However,we did in fact find that significant infall rates can still lead to a
decrease in the
$D$ abundance from the pre-solar value to today.

In our study, we have always chosen a primordial value of $D/H = 7.5 \times
10^{-5}$ corresponding to a baryon-to photon ratio of $\eta \simeq 3 \times
10^{-10}$.  We then selected solutions which yield a present deuterium
abundance
of $\approx 1.6 \times 10^{-5}$ in order to match the ISM measurement (Linsky
et al., 1992). In point of fact, we had solutions with both more and less
deuterium destruction. Thus, our solutions should not be viewed as being tied
to
our choice of $\eta$ or to our assumed primordial abundance of $D/H$.
Had we chosen the much higher value of $D/H \approx 2 \times 10^{-4}$
(Songaila et al. 1994;
Carswell et al. 1994) we would have still found solutions albeit many fewer of
them.

We have also found that adjusting the value of $m_{upp}$ to match the solar
$^{16}O$ abundance can have a dramatic effect on the evolution of $D$.  In
the case of the group IIa models,  $D$ is destroyed almost entirely by
lowering $m_{upp}$.  It is clear that such solutions exist as they require
nominally less $D$ destruction than those we have already found. Those models
which have exponential infall rates (groups Ic, IId) result in the opposite
problem.  Not enough $D$ is destroyed for these model types when lowering
$m_{upp}$.

Any number of our models could be considered good candidates for the
evolution of $D/H$.  Usually, those models which match observations and result
in a gas fraction of $\sigma \sim .1 - .2$ require too much infall onto the
disk.  Those models with a lower infall rate result in a $\sigma < .1$. (Bear
in mind however, our earlier remarks concerning the relationship between the
IMF
and $\sigma$ as well as the effect of our calculational scheme on $\sigma$.)
These models all assumed infall gas with a primordial
$D$ abundance.  Lowering the abundance of $D$ in the infall gas could result in
models which can better fit $\sigma$ without excess infall.

Our models resulted in the age of the galaxy in the range of $9.8-21.6$ Gyrs.
While nuclear
chronometers may not constrain
the age of the galaxy very tightly, they have proven to be a
useful tool in determining viable chemical histories for the reproduction of
$D$ observations. They give epochs with which to compare with the
observations within the models and eliminate those combinations of parameters
which can not reproduce the observed
ratios of nuclear chronometers.

\section{Acknowledgments}
 We would like to thank
Jean Audouze, Michel Cass\'{e}, Friedrich-Karl Thielemann, David Schramm and
Elisabeth Vangioni-Flam
 for very helpful discussions.
This work was supported in part by  DOE grant DE-FG02-94ER-40823.

\newpage
\beginapjbib

\bibitem Anders, E., \& Ebihara, M. 1982, Geochim. Cosmochim. Acta, 46,
2363

\bibitem Audouze, J., \& Tinsley, B. M. 1974, ApJ, 192, 487

\bibitem Carswell, R.F., Rauch, M., Weymann, R.J., Cooke, A.J. \&
Webb, J.K. 1994, MNRAS, 268, L1

\bibitem Clayton, D. D. 1984, ApJ, 285, 411

\bibitem Clayton, D. D. 1985, ApJ, 290, 428

\bibitem Cowen, J. J., Thielemann, F.-K., \& Truran, J. W. 1991, Annu. Rev.
Astron. Astrophys., 29, 447

\bibitem Cowen, J. J., Thielemann, F.-K., \& Truran, J. W. 1987, ApJ, 323,
543

\bibitem Delbourgo-Salvador, P., Gry, C., Malinie, G., \& Audouze, J. 1985,
A\& A, 150, 53

\bibitem Ferlet,  R. 1992, in {\it Astrochemistry of cosmic phenomena},
Ed. P.D. Singh (Kluwer, Netherlands), p. 85

\bibitem Geiss, J. 1993, in {\it Origin and Evolution of the Elements}
eds. N. Prantzos, E. Vangioni-Flam, and M. Cass\'{e} (Cambridge:Cambridge
University Press), p. 89

\bibitem Iben, I., \& Tutukov, A. 1984, ApJ Supp, 54, 335

\bibitem Larson, R.B. 1986, MNRAS, 218, 409

\bibitem Linsky, J. L., Brown, A., Gayley, K., Diplas, A., Savage, B. D.,
Ayres, T. R., Landsman, W., Shore, S. N., Heap, S. R. 1992, ApJ, 402, 694

\bibitem Linsky, J.L. 1994,  talk presented at the ESO/EIPC Workshop on
the Light Element Abundances

\bibitem Lynden-Bell, D. 1975, Vistas Astr., 19, 299

\bibitem Olive, K.A., \& Steigman, G. 1994, University of Mnnesota preprint
UMN-TH-1230/94.

\bibitem Olive, K.A., Theilemann, F.-K., \& Truran, J.W. 1987, ApJ, 313, 813

\bibitem Pagel, B. E. J., Simonson, E. A., Terlevich, R. J., \& Edmunds, M.
1992, MNRAS, 255, 325

\bibitem Pagel, B. E. J., \& Patchett, B. E. 1975, MNRAS, 172, 13

\bibitem Rana, N. C. 1991, Ann. Rev. Astron. Astrophys., 29, 129

\bibitem Reeves, H. 1991, A\& A, 244, 294

\bibitem Reeves, H. 1994, Rev. Mod. Phys. (in press)

\bibitem Scalo, J. M. 1986, Fund. Cos. Phys., 11, 1

\bibitem Skillman, E., \etal\ 1994b, ApJ Lett (in preparation)

\bibitem Songaila, A., Cowie, L.L., Hogan, C. \& Rugers, M. 1994
Nature, 368, 599

\bibitem Spite, F. \& Spite, M. 1993, in {\it Origin
 and Evolution of the Elements} eds. N. Prantzos,
E. Vangioni-Flam, and M. Cass\'{e}
(Cambridge:Cambridge University Press), p.201

\bibitem Steigman, G., \& Tosi, M. 1992, ApJ, 401, 150

\bibitem Steigman, G. 1994, Ohio State University preprint OSU-TA-7/94

\bibitem Tinsley, B.M. 1980, Fund. Cosm. Phys., 5,287

\bibitem Tosi, M. 1988, A\&A, 197, 33

\bibitem Twarog, B. A. 1980, ApJ, 242, 242

\bibitem Vangioini-Flam, E., \& Audouze, J. 1988, A\& A, 193, 81

\bibitem Vangioni-Flam, E., Olive, K.A., \& Prantzos, N. 1994,
ApJ, 427, 618 (VOP)

\bibitem Vidal-Madjar, A. 1991, Adv Space Res, 11, 97

\bibitem Vidal-Madjar, A., \& Gry, C. 1984, A\& A, 138, 285

\bibitem Woosley, S.E. 1993, Les Houches summer school on
Supernovae (to be published)
\endapjbib

\clearpage

\begin{table*}
\begin{center}
\begin{tabular}{crrrrrrrrrrr}           \hline \hline
Model   & SFR & IMF & $M_{low}$ & $M_{upp}$ & Infall Rate \\  \hline
Ia     &   $\nu M_G$  & $M^{-2.35}$ & .4 M\sun & 100 M\sun & $\lambda M_G$   \\
Ib     & "  & Tinsley   & .1 M\sun & 100 M\sun &    "           \\
Ica     & " & $M^{-2.35}$ & .4 M\sun & 100 M\sun &$\lambda e^{-t/\infty}$ \\
Icb     & " & " & " & " & $\lambda e^{-t/10}$ \\
Icc     & " & " & " & " & $\lambda e^{-t/8}$ \\
Icd     & " & " & " & " & $\lambda e^{-t/5}$ \\
Ice     & " & " & " & " & $\lambda e^{-t/3}$ \\
IIaa  &   $\nu e^{-t/\infty}$ & " & " & " & $\lambda M_G$   \\
IIab  &   $\nu e^{-t/10}$     & " & " & " & "           \\
IIac  &   $\nu e^{-t/8}$     & "  & " & " & "  \\
IIad  &   $\nu e^{-t/5}$     & "  & " & " & "  \\
IIae  &   $\nu e^{-t/3}$     & "  & " & " & "  \\
IIba   &   $\nu e^{-t/\infty}$     & Scalo & .1 M\sun & 100 M\sun & "  \\
IIbb   &   $\nu e^{-t/10}$     & " & " & " & "  \\
IIbc   &   $\nu e^{-t/5}$     & " & " & " & "  \\
IIbd   &   $\nu e^{-t/3}$     & " & " & " & "  \\
IIc & $\nu e^{-t/10}$ & $M^{-2.35}$ & .4 M\sun & 100 M\sun & $\lambda
e^{-t/10}$\\
IId & $\nu e^{-t/10}$ & Tinsley & .1 M\sun & 100 M\sun & $\lambda M_G$\\
III    &   $\nu {M_G}^2$       & $M^{-2.05}$ & .4 M\sun & 100 M\sun & "  \\
\hline
\end{tabular}
\end{center}
\caption{Summary of Model Parameters} \label{tbl-2}
\end{table*}

\clearpage

\beginfig

\figitim {\bf Figure 1a:} Relationship between the Total Mass of the galaxy
in units of initial disk mass and the initial star formation rate for group
Ia (filled circles), Icb
(filled boxes), and Icc (Filled triangles)  model solutions.

\figitim{\bf Figure 1b:} Same as figure (1a)
for models
IIaa (x's), IIac (open triangles), IIad (open boxes), and IIae (open
circles).

\figitim {\bf Figure 2:} Relationship between the Gas Mass Fraction
($\sigma$)
and the Total Mass of the galaxy in units of initial disk mass for groups
IIaa (x's), IIac (open triangles), IIad (open boxes), and IIae (open
circles) model solutions.

\figitim {\bf Figure 3a:} Relationship between the Initial Enrichment
($S_0$)
and the initial star formation rate for models Ia (filled circles), Icb
(filled boxes), and Icc (Filled triangles). The cutoff at $\nu = 1$
only represents the range of parameters tested.

\figitim{\bf Figure 3b:} Same as figure (3a) for models
IIaa (x's), IIac (open triangles), IIad (open boxes), and IIae (open
circles).

\figitim {\bf Figure 4a:} Relationship between the Initial Enrichment
($S_0$)
and the resulting age of the galaxy at the pre-solar epoch for models Ia
 (filled circles), Icb
(filled boxes), and Icc (Filled triangles).

\figitim{\bf Figure 4b:} Same as figure (4a) for models
IIaa (x's), IIac (open triangles), IIad (open boxes), and IIae (open
circles).

\figitim {\bf Figure 5a:}  Corresponding $D$ evolution for
the models in figure (5a) including the observations at the pre-solar epoch
(Geiss 1993) and today (Linsky et al. 1992).

\figitim {\bf Figure 5b:} Evolution of the gas fraction for
models
Ia (solid line), Ib (dotted line), Icb (short dashed line), and IIab (long
dashed line) which resulted in a pre-solar epoch of $9.6 Gyrs$.

\figitim {\bf Figure 6:} $D$ evolution for group Ia model assuming infall $D$
abundances of primordial (solid line), 50\% primordial (dotted line),
and no $D$ (dashed line).

\endfig

\end{document}